\documentclass[conference,compsoc]{IEEEtran}

\newcommand{\pactsourcepath}{sp2027/}

\usepackage{graphicx}
\graphicspath{{sp2027/}}
\usepackage{booktabs}
\usepackage{multirow}
\usepackage{tabularx}
\usepackage{array}

\usepackage{amsmath}
\usepackage{amssymb}
\usepackage{amsfonts}
\usepackage{dsfont}       

\usepackage{xurl}        
\usepackage{microtype}
\usepackage{xspace}

\usepackage{algorithm}
\usepackage{algpseudocode}

\usepackage{float}
\usepackage{xcolor}
\usepackage{comment}

\usepackage{pifont}
\newcommand{\cmark}{\ding{51}}

\newcolumntype{L}{>{\raggedright\arraybackslash}X}
\newcolumntype{P}[1]{>{\raggedright\arraybackslash}p{#1}}
\newcommand{\tablestyle}{\footnotesize\setlength{\tabcolsep}{5pt}\renewcommand{\arraystretch}{1.1}}
\newcommand{\rowgap}{\addlinespace[3pt]}
\newcommand{\thead}[1]{\textbf{#1}}

\makeatletter
\def\paragraph{\@startsection{paragraph}{4}{\z@}{0ex plus 0.1ex minus 0.1ex}%
{0ex}{\normalfont\normalsize\bfseries}}
\makeatother


\usepackage[hidelinks]{hyperref}

\newcommand{\name}{\textsc{PACT}\xspace}
\newcommand{\namenospace}{\textsc{PACT}}

\title{\namenospace: Making Privacy Governance Executable
at Data Interfaces}

\IfFileExists{authors.tex}{\author{\IEEEauthorblockN{Sangshin Park, Nazmus Shakib Sayom, Manish Parashar, and Luis Garcia}
\IEEEauthorblockA{Kahlert School of Computing, University of Utah\\
Salt Lake City, Utah, USA}}
}{}

\begin{document}

\maketitle


\begin{abstract}


Privacy evaluations can reveal sensitive information without
establishing whether an observed exposure violates an applicable
governance requirement. Existing approaches often separate empirical
privacy testing from the documented commitments governing data
collection, release, and use. We present \name, a provenance-aware
framework that connects documented privacy governance to executable
interface checks and evidence-bounded privacy evaluation. \name\
represents governance statements with explicit provenance and scope,
distinguishes documented requirements from evaluator assumptions,
and interprets privacy observations against applicable commitments
while preserving uncertainty.

We evaluate \name\ across an IRB-governed sensing study,
clinical-data releases, anonymized urban video, and operational
research-software interfaces. A protocol-derived validator detects
all eight injected violations of explicit release exclusions,
with no false positives across nine checks on a clean package.
Retrospective evaluations reveal substantial inferential exposure
even when documented privacy mechanisms operate as specified.
For example, MIMIC representations support over 97\% candidate
retrieval under the evaluated adversary despite independent
identifier regeneration, without establishing patient identity
or governance noncompliance.

We evaluate whether governance interpretations remain stable
across independently reconstructed and automatically extracted
evidence representations, finding substantial agreement alongside
representation-dependent differences that limit claims of
correctness. \name\ thus connects empirical privacy measurements
to auditable, evidence-bounded interpretations of documented
governance requirements.

\end{abstract}



\section{Introduction}

MIMIC-III and MIMIC-IV, two versions of the Medical Information Mart
for Intensive Care (MIMIC), are widely used critical-care datasets
distributed separately through PhysioNet under credentialed access
\cite{PhysioNet-mimiciii-1.4,PhysioNet-mimiciv-3.1}. Their documentation
states that the underlying patient populations overlap during
2008--2012 and that identifiers are regenerated independently for each
release. Records from the two releases can nevertheless remain
associable through recurring released content. Such an association does not, by itself, establish
patient identity, a de-identification failure, or a violation of a
documented commitment: the overlap is documented, and the identifier
mechanism works as described. Its meaning depends on which documented
statements apply to the measured property, and at what scope.

Treating successful inference as a violation presumes a commitment that
covers the measured property; treating a documented mechanism, such as
identifier regeneration or face blurring, as a guarantee confuses
applying a procedure with preventing inference. Distinguishing these
cases requires examining each source statement's semantic role and
scope, separately from the targets, adversaries, metrics, and thresholds
an evaluator introduces to make a question testable.

Existing work addresses parts of this problem in isolation. Empirical
privacy evaluation measures what an interface still reveals under
evaluator-chosen targets and thresholds
\cite{apthorpe2019smarter,noorbakhsh2024inf2guard,paul2026sensor};
policy analysis compares privacy statements with observed behavior
\cite{andow2020policheck,cui2023poligraph,trimananda2022ovrseen}; and
compliance work checks explicit accesses, dataflows, or programs against
computable requirements
\cite{byun2005purpose,park2004ucon,sen2014bootstrapping,wang2022privguard,lawson2021duo,cabili2021duos}.
These approaches do not jointly address the task of relating
interface-specific observations to applicable, source-supported
governance propositions while keeping evaluator-defined assumptions
distinct from documented requirements.

We present \name\ (Provenance-Aware Audit and Contract Translator), a
provenance-aware framework that makes this connection explicit. \name\
records each reviewed governance statement as a source-linked
proposition with a semantic role (outcome commitment, mechanism claim,
access/recipient condition, purpose restriction, acknowledged risk, or
contextual fact), an origin, an evidence state, and a scope, and binds
applicable propositions to a versioned interface as an interface
contract; the evaluator's target, adversary, metric, and controls are
recorded separately. After evaluation, \name\ asks whether the reviewed
evidence grounds the question at all, and whether any supported
proposition applies to the measured property. Only an applicable
proposition can make an observation \emph{Consistent} or
\emph{Contradictory}; a grounded observation with no applicable
proposition has \emph{No Applicable Commitment}, and insufficient
evidence remains \emph{Unresolved}. The executable package begins from
source-linked records, built in our study by coding agents under
investigator supervision, and does not perform unrestricted
natural-language governance extraction.

We evaluate \name\ on three governance-grounded datasets and on
operational research infrastructure. In STMicroelectronics Human
Activity Recognition (STM-HAR), a
wearable-sensing research dataset under an active Institutional Review
Board (IRB) protocol, a
protocol-derived validator detects all eight injected violations of
explicit exclusions, the only \emph{Contradictory} relations in our
evaluation. In MIMIC, a ranking adversary places the recurring
counterpart first for more than 97\% of records in
the recurring populations; recurrence and candidate association have
\emph{No Applicable Commitment}, while the identifier and CareVue
exclusion mechanisms are \emph{Consistent}. In StreetAware, whose
third-party source documents face blurring \cite{piadyk2023streetaware},
blurring is applied as documented, while within-recording association
through blurred face crops and through body and motion cues has
\emph{No Applicable Commitment}. As a consistency check rather than
ground-truth validation, six blinded agent reconstructions of the
evidence agreed on the six decisions central to these interpretations. At
operational interfaces, \name\ keeps software capability, deployment,
authorization, and enforcement apart, including for a client-side signed
upload check.

These distinctions let researchers report what an attack demonstrates
without overstating it, and require stating which adversary a clause
addresses: MIMIC's prohibition on re-identification constrains what
credentialed recipients may do, not what the released representations
permit, just as misuse of license-plate, clinical, and home-video
systems by authorized users shows that technical access and authorized
use diverge \cite{flockMilwaukee2026,hernandezRoman2025,ftcRing2023}.
Neither \emph{Consistent} nor \emph{No Applicable Commitment} certifies
privacy safety, and \name\ does not itself enforce governance or
determine legal compliance.

As a secondary analysis, we test how sensitive these interpretations are
to the evidence representation, comparing PACT's manually reviewed
records, originally constructed with LLM coding-agent assistance and
serving as comparison anchors rather than ground truth, with MIMIC and
StreetAware representations produced by an automated LLM-based
extraction pipeline under the unchanged relation procedure. Downstream
relations largely agreed, while intermediate grounding and applicability
judgments differed; agreement does not establish correctness, and
automated extraction is not a proven replacement for PACT's records.

\textbf{Contributions.}
(1) A provenance-preserving evidence representation in which each
governance statement retains its source span, semantic role, origin,
evidence state, and scope, bound to versioned interfaces as contracts
that drive executable interface checks while keeping evaluator-defined
operationalizations separate (Section~\ref{sec:interface_contract}).
(2) A grounding- and applicability-based procedure that relates each
observation to specific applicable propositions as one of the four
relations, under which successful inference alone cannot be
\emph{Contradictory} (Section~\ref{sec:cr_adversary_formalization}).
(3) An evaluation on STM-HAR, MIMIC-III/IV, and StreetAware showing that
substantial exposure coexists with mechanisms operating as documented
and with \emph{No Applicable Commitment}, that the observed
\emph{Contradictory} relations arise only from explicit applicable
exclusions, and that central decisions were stable across six blinded
agent reconstructions (Section~\ref{sec:evaluation}), extended to
operational research-data interfaces (Section~\ref{sec:eval_system}).
(4) An evaluation of LLM-extracted governance representations as
alternative inputs to the unchanged procedure, tracing how extracted
roles, applicability, and evidence states shape downstream relations and
their routes (Sections~\ref{sec:eval_spectra_rq1}
and~\ref{sec:eval_spectra_rq2}).

\section{Background and Related Work}
\label{sec:background}

\begin{figure}[t]
    \centering
    \includegraphics[width=0.8\columnwidth]{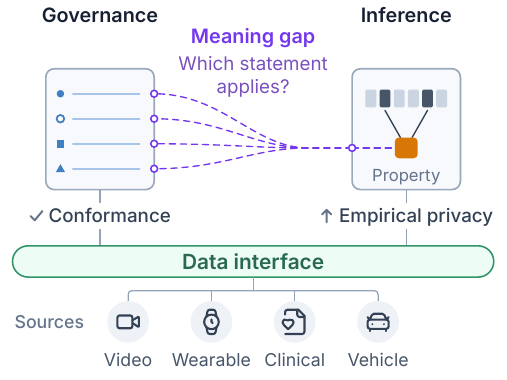}
    \caption{Complementary privacy questions around a data interface.
    Solid edges are questions addressed by existing work: governance
    (permitted or required use), conformance (whether explicit behavior
    satisfies it), and empirical evaluation (what remains inferable). The
    dashed edge is the question \namenospace\ addresses: what an
    observed inference means relative to the governing evidence. Icons
    illustrate data modalities, not evaluation cases.}
    \label{fig:related_work_landscape}
\end{figure}

Privacy research asks several related questions around a common data
interface (Figure~\ref{fig:related_work_landscape}): governance and
policy systems represent which accesses, purposes, or information flows
are permitted; conformance systems test whether explicit software
behavior satisfies those requirements; and empirical privacy evaluation
measures what remains inferable from the resulting representation.
\name\ addresses the interpretive question on the dashed edge.

\subsection{Governance, Policy, and Conformance}
\label{subsec:bg_governance}

Purpose- and usage-control models constrain purpose, context,
obligations, and post-access conditions \cite{byun2005purpose,park2004ucon},
and contextual-integrity formalizations represent actors, recipients,
and transmission norms \cite{barth2006contextual}. LEGALEASE/GROK and
PrivGuard check dataflows or analysis programs against machine-readable
policies \cite{sen2014bootstrapping,wang2022privguard}; DUO and DUOS
encode consent-derived restrictions for biomedical access
\cite{lawson2021duo,cabili2021duos}; and PoliCheck, PoliGraph, OVRseen,
and Paralegal compare policy assertions with application behavior,
network traffic, or implementations
\cite{andow2020policheck,cui2023poligraph,trimananda2022ovrseen,
adam2025paralegal}. These approaches ask whether explicit accesses,
purposes, dataflows, or programs conform to represented policy; \name\
relates an empirical property of the released interface back to the
evidence without assuming that the property was itself a commitment.

\subsection{Empirical Privacy at Data Interfaces}
\label{subsec:bg_privacy}

Anonymization, redaction, aggregation, and representation minimization
can reduce direct exposure without eliminating privacy-relevant
semantics \cite{9973552,selective,cangialosi2021prividpracticalprivacypreservingvideo}.
Inference succeeds from encrypted traffic, learned representations, and
low-dimensional sensor outputs
\cite{apthorpe2019smarter,noorbakhsh2024inf2guard,paul2026sensor}, and
independently prepared releases without shared identifiers can remain
associable through recurring content \cite{11211185}. Sensor Privacy as
a Spectrum formalizes empirical privacy relative to an observable
representation, sensitive target, adversarial context, and leakage
measure \cite{paul2026sensor}. Such targets, adversaries, metrics, and
thresholds are empirical operationalizations; the measurement alone does
not determine whether the governing evidence makes a corresponding
commitment.

\subsection{Connecting Governance Conformance and Empirical Privacy}
\label{subsec:bg_gap}

Neither conformance nor successful inference says what an observed
inference means relative to the governing evidence; the answer depends
on the semantic role and scope of the source statement. A
\emph{mechanism claim} that an anonymization procedure was applied is
not an \emph{outcome commitment} that information is non-inferable; a
\emph{purpose restriction} or \emph{access/recipient condition}
constrains how or by whom information may be used rather than whether it
is inferable; and observing an \emph{acknowledged risk} instantiates
what the source already recognizes rather than contradicting it. \name\
provides this provenance-preserving bridge rather than a new policy
language; it keeps missing or conflicting evidence explicit and does not
automate legal interpretation or establish general privacy safety.

\section{PACT Framework}
\label{sec:framework}

\subsection{Overview and Design Principle}
\label{sec:framework_overview}

\begin{figure*}[t]
\centering
\includegraphics[width=\textwidth]{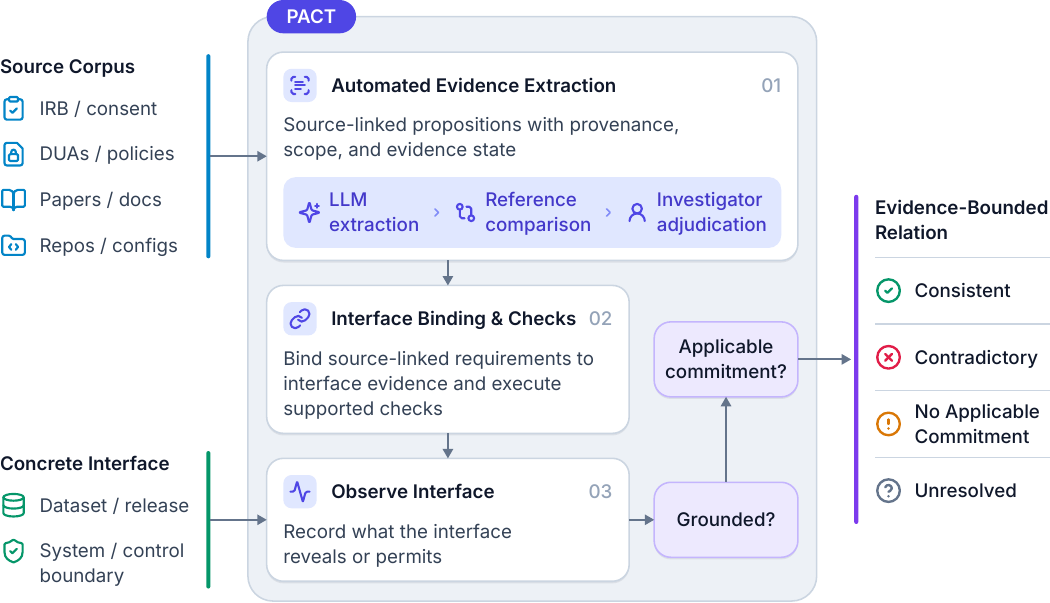}
\caption{\namenospace\ separates automated extraction of source-linked
evidence from interface binding and checks and from evidence-bounded
interpretation. Extracted records are compared with a reviewed
reference and adjudicated by an investigator; agreement does not
establish correctness. Governance evidence may ground an empirical
question, but only an applicable supported proposition can yield a
\emph{Consistent} or \emph{Contradictory} relation.}
\label{fig:framework_pipeline}
\end{figure*}

\name\ connects source-linked governance evidence to concrete data and
software interfaces without promoting evaluator choices into governance
requirements. Its pipeline (Figure~\ref{fig:framework_pipeline})
represents evidence in a typed, provenance-linked contract, binds and
operationalizes it at a concrete interface, evaluates that interface,
and relates the observation only to applicable supported propositions.

\paragraph{Core concepts}
A \emph{governance source} is a document with authority over a dataset
or deployment, such as an IRB protocol, data-use agreement, release
documentation, or dataset publication. Each reviewed statement becomes a
\emph{source-linked proposition}, a normalized claim with its supporting
source span, whose \emph{semantic role} is an outcome commitment (some
information will not be inferable), a mechanism claim (a procedure is
applied), an access/recipient condition, a purpose restriction, an
acknowledged risk, or a contextual fact. An \emph{interface contract} is
the subset of propositions applying to one versioned interface. Testing
a question there usually requires choices the source does not make (a
target, adversary, metric, and controls): the \emph{evaluator-defined
operationalization}, whose result is the \emph{empirical observation}.
Interpretation asks whether the evidence justifies asking the question
(\emph{grounding}, $G$) and whether any supported proposition covers the
measured property (the \emph{applicable commitments}, $Q$); only
propositions in $Q$ can make an observation \emph{Consistent} or
\emph{Contradictory} (Section~\ref{sec:interface_contract}). Record
construction is separate: Sections~\ref{sec:eval_spectra_rq1}
and~\ref{sec:eval_spectra_rq2} evaluate automated LLM-based governance
extraction as an alternative construction path with the relation
procedure unchanged.

\subsection{Deployment, Governance, and Interface Model}
\label{sec:system_model}

\name\ evaluates a concrete \emph{deployment context}: the applicable
governance evidence, the exposed interface, and the parties able to
observe or invoke it. A \emph{PACT interface} is the versioned data,
software, or control boundary at which an observation is exposed or an
operation is accepted, such as a dataset representation, API, model
output, uploader, converter, or configuration boundary. Recipients and
evaluator-defined adversaries are evidenced separately.

\label{sec:released_interfaces}

For a static dataset release with transformation $\tau$,
$O_\tau(x)=\tau(x)$ denotes the representation exposed for record $x$,
so different transformations, feature sets, or sensor subsets define
distinct interfaces.

\paragraph{Running example}
In MIMIC-III and MIMIC-IV
\cite{PhysioNet-mimiciii-1.4,PhysioNet-mimiciv-3.1}, five elements must
be kept apart: (i)~the documented 2008--2012 population overlap, a
\emph{contextual fact}; (ii)~independent identifier regeneration per
release, a \emph{mechanism claim}; (iii)~credentialed access without
re-identification attempts, an \emph{access/recipient condition} on how
data may be used, not on what released representations permit;
(iv)~exact cross-release \emph{representation recurrence} of an
admission's released summary, an \emph{empirical observation}; and
(v)~\emph{candidate association}, ranking target-release records for a
source-release record, an \emph{evaluator-defined operationalization}
that no source statement requires or prohibits. The overlap grounds the
association question, but the documented non-linkage concerns
identifiers, and no supported proposition commits to non-association of
the released summaries, so recurrence and association have \emph{No Applicable
Commitment}, whereas observations about the identifiers are
\emph{Consistent} with the mechanism claim
(Sections~\ref{sec:threat_construction} and~\ref{sec:eval_mimic}).

\subsection{From Contract to Interpretation}
\label{sec:framework_handoff}
\label{sec:framework_threat}

Binding an interface contract (Section~\ref{sec:interface_contract})
cannot strengthen a mechanism into an outcome guarantee, convert context
into a prohibition, or resolve missing evidence by assumption.

A \emph{governance-derived threat} is an empirical question bound to
supported evidence. The source need not specify the experiment; the
evaluator-supplied target, objective, auxiliary knowledge, metric,
threshold, and controls remain operationalization rather than
governance.

PACT evaluates the bound interface under explicit assumptions and
records the observation with the evaluated version or configuration, in
retrospective audits and in prospective validation at release, upload,
conversion, configuration, and import boundaries.

Interpretation uses four governance relations
(Section~\ref{sec:cr_adversary_formalization}): \emph{Consistent}
(compatible with an applicable proposition), \emph{Contradictory}
(directly conflicting with one), \emph{No Applicable Commitment} (a
grounded observation that no supported proposition constrains), and
\emph{Unresolved} (insufficient, ambiguous, conflicting, or withheld
evidence). \textsc{ObservationOnly} is not a fifth relation but an
ungrounded empirical result, and no relation establishes privacy safety
or general compliance.

\section{Evidence Model and Implementation}
\label{sec:interface_contract}

\subsection{Evidence Records and Interface Contracts}
\label{sec:contract_representation}

\name\ describes source-linked governance evidence as typed,
provenance-linked \emph{evidence records} rather than a fixed set of
policy slots. For a source-linked proposition, we write

\begin{equation}
e =
\langle
s,\; q,\; r,\; o,\; \eta,\; \sigma
\rangle,
\label{eq:evidence_record}
\end{equation}

where $s$ identifies the originating source and span, $q$ is the
normalized proposition, $r$ its semantic role
(Section~\ref{sec:framework_overview}), $o$ its origin, $\eta$ its
evidence state, and $\sigma$ its applicable scope. Roles are values,
not mandatory fields: a source may yield several records of one role and
none of another, a mechanism does not imply an outcome commitment, and a
contextual fact becomes an acknowledged risk only if the source supports
that reading. The origin $o$ is \emph{source-direct} or
\emph{interpretation-derived}; independently, $\eta$ is
\emph{supported}, \emph{unresolved}, \emph{conflicting},
\emph{withheld}, \emph{not applicable}, or \emph{not assessed}, so an
interpretation-derived proposition is never presented as source-direct
and missing or conflicting evidence stays explicit. Scope $\sigma$ (the
population, recipient, purpose, release, deployment, operation, or
temporal context) is part of the record, not an assumption added at
interpretation time.

In the running example, the identifier and overlap statements become
two records with different roles:
$e_{\mathrm{id}}=\langle s_{\mathrm{doc}},\,
q_{\mathrm{id}},\,\textit{mechanism claim},\,\textit{source-direct},\,
\textit{supported},\,\sigma_{\mathrm{id}}\rangle$, where $q_{\mathrm{id}}$
states that identifiers are generated independently for each release
and $\sigma_{\mathrm{id}}$ covers the released identifier fields; and
$e_{\mathrm{ov}}=\langle s_{\mathrm{doc}},\,q_{\mathrm{ov}},\,
\textit{contextual fact},\,\textit{source-direct},\,\textit{supported},\,
\sigma_{\mathrm{ov}}\rangle$, where $q_{\mathrm{ov}}$ states that the
source populations overlap during 2008--2012 and $\sigma_{\mathrm{ov}}$
covers the two releases' patient populations.

The tuple is a semantic model, not the runtime JSON schema or a claim
that each dimension is produced automatically from source text; the
executable package consumes records produced before runtime
(Section~\ref{sec:compiler}). An \emph{interface contract} is the subset
of evidence records bound to a concrete interface and deployment
context:

\begin{equation}
C_{I,d,v}
=
\mathrm{Bind}(E,I,d,v),
\label{eq:interface_contract}
\end{equation}

where $E$ is the source-linked evidence set, $I$ the evaluated
interface, $d$ its release or deployment context, and $v$ the relevant
version or configuration. Binding does not change the propositions in
$E$; it determines which records and scopes apply. Evaluator-defined
elements and the resulting observations and relations are linked to the
contract (Section~\ref{sec:threat_construction}) but never written into
$C_{I,d,v}$ as governance propositions.

\subsection{Implementation and Execution Boundary}
\label{sec:compiler}

\name\ is implemented as a Python package with eleven JSON Schemas for contracts,
deployments, evidence records, and related artifacts. Deployment artifacts bind the
applicable contract and governance artifacts by identifier, version,
and digest; the validator checks these bindings, evidence states,
interface constraints, and review triggers, and emits \textsc{PASS},
\textsc{WARNING}, \textsc{FAIL}, or \textsc{REQUIRES\_REVIEW} findings
that retain their governance-artifact provenance. Governance relations
remain interpretation-layer outputs, not validator status codes. For
example, the validator
on the public StreetAware contract/deployment pair returns
\textsc{REQUIRES\_REVIEW} for retention and purpose because the source
specifies neither, rather than filling those terms by default.

\subsection{Operationalization and Grounding}
\label{sec:threat_construction}
Threat construction binds an evaluator-defined question to source-linked
evidence: governance may support a target directly, through an outcome
commitment or documented concern, or ground evaluation of a mechanism,
access condition, purpose, or context without implying an outcome
guarantee.


\label{sec:policy_targets}
We represent the evaluator-defined choices as an operationalization

\begin{equation}
u =
\langle
T,\; A,\; M,\; K
\rangle,
\label{eq:operationalization}
\end{equation}

where $T$ is the evaluation target, $A$ the adversarial or operational
objective, $M$ the empirical measure or validation rule, and $K$ the
auxiliary assumptions and controls; none is a proposition in
$C_{I,d,v}$. Before $u$ can receive a governance interpretation, it must
be grounded in the contract:

\begin{equation}
G(u,C_{I,d,v}) =
\begin{cases}
1, & \exists e \in C_{I,d,v}\ \text{s.t.}\ 
      \mathrm{Trace}(u,e)=1,\\
0, & \text{otherwise},
\end{cases}
\label{eq:grounding}
\end{equation}

where $\mathrm{Trace}(u,e)$ records an explicit connection between $u$
and the protected concern, mechanism or its stated purpose, access
condition, purpose restriction, acknowledged risk, or context in record
$e$. Evaluating interface $I$ under $u$ yields an observation
$z=\mathrm{Eval}(I,u)$, and evidence that merely \emph{grounds} the
question is distinguished from an applicable proposition about the
property observed:

\begin{equation}
\begin{aligned}
Q(C_{I,d,v},u)
= \{\,q_e \mid{} & e\in C_{I,d,v},
\ \mathrm{Applies}(e,u)=1,\\
& \eta_e=\mathrm{supported}\,\}
\end{aligned}
\label{eq:applicable_props}
\end{equation}

is the set of supported propositions whose recorded scope bears on the
property evaluated by $u$. An operationalization may thus have $G=1$
while $Q=\emptyset$: evidence can justify asking a question without
asserting what its answer must be. Contextual facts and
acknowledged risks may ground a question without becoming commitments
that an observation can contradict. A mechanism claim enters $Q$ only
for an observation about whether that mechanism is applied or behaves as
stated; outcome commitments, access/recipient conditions, and purpose
restrictions enter $Q$ only when the observation tests the corresponding
scoped outcome or restriction. Mere compatibility with contextual
evidence does not make an observation \emph{Consistent}.

In the running example, cross-release candidate association is
$u=\langle T,A,M,K\rangle$ with $T$ the target-release counterpart of a
source-release record, $A$ a ranking adversary with credentialed access
to both releases, $M$ Top-1 retrieval, and $K$ no cross-release
identifier mapping plus matched null controls. The overlap record gives
$\mathrm{Trace}(u,e_{\mathrm{ov}})=1$, so $G=1$. Neither
$e_{\mathrm{ov}}$, a contextual fact, nor $e_{\mathrm{id}}$, scoped to
the identifier fields, applies to non-identifier representations, and
the credentialed-access condition constrains recipient conduct rather
than the measured property; hence $Q=\emptyset$. An operationalization
that checks whether the released identifiers provide a cross-release
key instead satisfies $\mathrm{Applies}(e_{\mathrm{id}},\cdot)=1$ and
can be \emph{Consistent} with $e_{\mathrm{id}}$.

\paragraph{Adversary and controls}
\label{sec:capabilities}
The adversary observes $O_\tau$ with the auxiliary information of the
evaluation setting (public documentation, pretrained models, or other
released datasets), recorded as an evaluation choice rather than
attributed to the source. Where several model families are evaluated,
\name\ reports the strongest supported result; prior observations of a
target record are included only when the evaluation model justifies
them; and case-specific controls (permutation or matched nulls,
cross-release comparisons, ablations) provide the empirical reference.

\subsection{Evidence-Bounded Governance Relations}
\label{sec:cr_adversary_formalization}
PACT assigns governance meaning only after establishing whether an
evaluation is grounded, whether an applicable proposition exists, and
whether its supporting evidence is sufficient. With $G$ and $Q$ as in
Section~\ref{sec:policy_targets}, the relation procedure for $u$ and
observation $z$ is

\begin{equation}
\mathcal{R}(C,u,z)=
\begin{cases}
\textsc{ObservationOnly}, & G(u,C)=0,\\
\textsc{Unresolved}, &
  \substack{\mathrm{Evidence}(C,u)\\ \text{is insufficient}},\\
\substack{\text{\scshape NoApplicable}\\
          \text{\scshape Commitment}}, &
  Q(C,u)=\emptyset,\\
\{\rho(q,z)\}_{q\in Q(C,u)}, & \text{otherwise}.
\end{cases}
\label{eq:governance_relation}
\end{equation}

where the final relation is evaluated separately for each applicable
source-supported proposition $q$:

\begin{equation}
\rho(q,z)=
\begin{cases}
\textsc{Contradictory}, &
  \substack{z\text{ directly conflicts}\\ \text{with }q},\\
\textsc{Consistent}, &
  \substack{z\text{ is compatible}\\ \text{with }q},\\
\textsc{Unresolved}, &
  \substack{\text{the relation cannot}\\ \text{be established}.}
\end{cases}
\label{eq:proposition_relation}
\end{equation}

Insufficient evidence includes unresolved, conflicting, or withheld
evidence needed to establish applicability or interpret the relation.
\textsc{Consistent} means only that the observation does not contradict
the applicable proposition, not compliance, privacy safety, or
effectiveness beyond the tested property. Because the procedure checks
grounding, evidence sufficiency, and applicability before assigning a
proposition-level relation, successful inference cannot be
\textsc{Contradictory} without an applicable supported proposition for
the measured property. Because relations
are proposition specific, one interface may be \textsc{Consistent} with
a documented mechanism and have \textsc{NoApplicableCommitment} with
respect to a stronger guarantee the source never states.

\section{Evaluation}
\label{sec:evaluation}

\textbf{RQ1 (Evidence preservation and interpretation stability):} can
PACT preserve provenance, role, scope, and evidence state through
binding, and are downstream boundaries stable under independent
reconstruction of the evidence?
\textbf{RQ2 (Evidence-bounded interpretation):} can source-linked
evidence bound empirical evaluation and distinguish governance meanings
of observed exposure?
\textbf{RQ3 (Prospective validation):} can the representation drive
checks before release or deployment?
\textbf{RQ4 (Operational interfaces):} can PACT keep capability,
deployment, authorization, and enforcement apart at software boundaries?

\subsection{Cases and Evidence}
\label{sec:eval_overview}

Each case is used only where its evidence supports the claim. STM-HAR
(RQ1--RQ3) is the only case whose governance, an active IRB protocol,
predates release, so it supplies the prospective
and positive-control \emph{Contradictory} evidence. MIMIC (RQ1--RQ2)
tests retrospective cross-release exposure under credentialed access,
with a contract reconstructed from PhysioNet documentation
\cite{PhysioNet-mimiciii-1.4,PhysioNet-mimiciv-3.1}; its
independently curated CareVue subset, which excludes patients with a
MIMIC-IV hospitalization, serves as a counterfactual. StreetAware
(RQ1--RQ2) tests mechanism-scoped anonymization documented entirely by
a third party \cite{piadyk2023streetaware}. CAPTURE, a multimodal
research-data collection project, and Brain Behavior Quantification and
Synchronization (BBQS) test operational boundaries (RQ1, RQ4; BBQS also
RQ3) with deployment-specific,
often absent or unresolved governance evidence.

\subsection{RQ1: Evidence Preservation and Interpretation Stability}
\label{sec:eval_rq1}

PACT keeps documented mechanisms, contextual facts, and evaluator-defined
questions from becoming stronger governance propositions. STM-HAR's 23
recorded governance items include 14 source-direct and 9
interpretation-derived propositions; independently, five carry an
unresolved evidence state because the protocol does not specify the
relevant property. No MIMIC statement commits to cross-release
non-association beyond identifiers, and no StreetAware statement to
facial non-inferability, so identifier
regeneration, the overlap, and face anonymization remain mechanism or
context records and the association targets stay evaluator-defined;
STM-HAR's targets operationalize its acknowledged risk; and CAPTURE and
BBQS keep capability, deployment, authorization, and enforcement
evidence apart.

To test whether these boundaries merely reproduced author choices, six
blinded agent sessions (three per model family) repeated evidence
construction over identical fixed copies of the MIMIC paper, the CareVue
subset page, and the StreetAware paper, with the same codebook but
without prior contracts or classifications, other sessions'
outputs, or empirical results. Each session first built source-linked
records (Phase~A), which were frozen before the observations were
revealed, and then assigned relations from them (Phase~B). All six agreed on the six central decisions:
identifier regeneration and StreetAware blurring were \emph{Consistent}
mechanism observations, while MIMIC non-identifier recurrence, $>97\%$
retrieval on the established recurrence populations, blurred-face
association, and facial identity-nonrecovery tests had \emph{No
Applicable Commitment}. Variation concentrated in context-only grounding
and extraction completeness: overlap/CareVue relations varied, and some
sessions omitted a StreetAware tracking-context record and returned
\textsc{ObservationOnly} rather than backfilling Phase~A. This tests
procedural stability under a fixed corpus and codebook, not human
inter-rater reliability or the codebook's normative correctness.

\paragraph{Replication with LLM-extracted representations}
\label{sec:eval_spectra_rq1}
Do the six central decisions depend on how the evidence records were
constructed? This replication and its counterpart in
Section~\ref{sec:eval_spectra_rq2} compare PACT's manually reviewed
representations with governance representations produced by automated
LLM-based extraction. PACT's manually reviewed records were not
written manually: they were built with LLM coding agents under PACT's
codebook and then manually reviewed, and in this replication each
session receives the Phase~A records of one of the six blinded
ChatGPT/Gemini sessions above. This manually reviewed baseline
serves as a comparison anchor rather than independently established
ground truth, and the comparison tests how sensitive PACT's
interpretations are to the evidence representation, with replacement
assessed only by the stricter pre-specified rule below. We repeated the
reconstruction step with three Phase~A inputs, never combined:
\emph{Manually Reviewed};
\emph{DeepSeek}, 213 records extracted from the MIMIC and StreetAware
papers by an automated, multi-stage LLM governance extraction pipeline,
developed separately from PACT's relation procedure, using
DeepSeek-v4.1-Flash (extraction r-ds); and \emph{Qwen}, 223 records from
a separate run of the same pipeline with Qwen3.8-27B (r-qw). The two
runs share the pipeline's decision and review models; each extracted
record carries a proposition, semantic role, scope, and source span and
is converted into a Phase~A record. The eleven reconstruction questions
and the Phase~B protocol were frozen before any session ran. The corpora
differ: the pipeline used publisher XML of the two papers, with the
MIMIC-III article in place of the CareVue page that the blinded
sessions' PDF corpus included, so the one CareVue-dependent question is
reported only descriptively.

All three conditions used one model, Claude Sonnet~5.5 at reasoning
effort \emph{high}, in six fresh Claude Code CLI sessions each (18 valid
sessions in total), without tools, memory, or access to other outputs,
and with the hash-verified frozen input as the only input. These are
repeated draws from one model, not six model families, and supplement
rather than replace the ChatGPT/Gemini result above, in which the blinded
sessions preserved 36/36 central decision-instances with 6/11 unanimous
questions; results from the two model settings are not directly
comparable.

Table~\ref{tab:spectra_rq1} reports preservation of the six central
decisions over decision-instances (sessions~$\times$~6), question-level
unanimity, and citation validity (the share of cited identifiers that
resolve to a record in the session's own input); Qwen values were
checked only against author-held files. Both LLM-extracted
inputs preserved the central decisions at least as well as the manually
reviewed records under this model (71/72 decision-instances, against 30/36). The
counts are descriptive: they measure preservation of the blinded
sessions' decisions, not accuracy, and do not show that LLM-extracted
records cause better reconstruction, because the conditions differ in
more than the extraction method. Sessions of the Manually Reviewed condition received
different inputs (three of 41 to 52 records, three of 5), whereas every session of
an LLM-extraction condition received the same 213 or 223 records, which
favors within-condition unanimity and gives far greater coverage; the
Manually Reviewed condition's three StreetAware tracking-context questions split
exactly by input group. In three sessions of that condition, the model returned
\emph{Unresolved} for non-identifier recurrence and the $>97\%$
retrieval result, treating a manually reviewed record stating that the releases
``cannot be linked'' as ambiguous in scope, which Phase~A had left
unresolved. The only deviation with LLM-extracted records is one
DeepSeek session returning \emph{Consistent} for facial identity
non-recovery, at medium confidence and explicitly as mere compatibility.
No session returned \emph{Contradictory}.

\begin{table}[t]
\centering
\caption{RQ1 replication results across manually reviewed and
LLM-extracted governance records.}
\label{tab:spectra_rq1}
\small
\setlength{\tabcolsep}{3pt}
\begin{tabularx}{\columnwidth}{@{}>{\raggedright\arraybackslash}X ccc@{}}
\toprule
Measure & \begin{tabular}[b]{@{}c@{}}Manually\\Reviewed\end{tabular} & DeepSeek & Qwen \\
\midrule
Central decisions preserved (instances) & 30/36 & 35/36 & 36/36 \\
Sessions preserving all six & 3/6 & 5/6 & 6/6 \\
Unanimous questions (of 11) & 4 & 8 & 9 \\
\emph{Contradictory} relations & 0 & 0 & 0 \\
Cited identifiers resolving & 216/216 & 399/399 & 418/420$^{\ast}$ \\
\bottomrule
\end{tabularx}

\vspace{2pt}\footnotesize
$^{\ast}$One identifier cited twice with a mistyped dataset prefix.
\end{table}

The pre-specified rule for replacing the manually reviewed records with
LLM-extracted ones retains the manually reviewed records. The rule was fixed in the analysis
script, whose hash was recorded before any session output existed, but
was not registered externally and was written after PACT's primary
results were known. It requires (i)~unanimous preservation of all six
central decisions in \emph{both} LLM-extraction conditions, which the
single DeepSeek deviation fails and which alone determines the verdict;
and (ii)~modal agreement with the Manually Reviewed condition on the primary
questions, which also fails because that condition itself departs from
the blinded sessions' decisions under this model: its modes differ from
the LLM-extraction modes on three primary StreetAware tracking-context
questions and three secondary MIMIC questions. Separately, two central
MIMIC questions tie 3--3 in the Manually Reviewed condition between \emph{No
Applicable Commitment} and \emph{Unresolved}, so it has no single mode
for them. Automated extraction is therefore not validated as a
replacement; the verdict concerns replacing the manually reviewed records in
this evaluation, not the quality of LLM-extracted records in general.

\subsection{RQ2: Evidence-Bounded Interpretation}
\label{sec:eval_grounded}

Across the evaluated cases, governance interpretation depends on
applicable, source-supported propositions rather than inference
success alone. Substantial inferential exposure can coexist with
documented mechanisms operating as specified, while the observed
\emph{Contradictory} relations arise from violations of explicit,
applicable exclusions.
Table~\ref{tab:evidence_chains} traces representative observations from
governance source to relation, naming the branch of the relation
procedure (Eqs.~\ref{eq:governance_relation}
and~\ref{eq:proposition_relation}) that yields each.

\begin{table*}[t]
\centering\small
\caption{Representative evidence chains: governance source and
proposition (semantic role), evaluated interface or test, empirical
observation, and the resulting relation with the decision-procedure
branch that yields it. $G1$--$G3$ denote each case's own candidate
representations (Sections~\ref{sec:eval_stmhar} and~\ref{sec:eval_mimic}).}
\label{tab:evidence_chains}
\begin{tabularx}{\textwidth}{@{}
  >{\raggedright\arraybackslash}p{1.5cm}
  >{\raggedright\arraybackslash}X
  >{\raggedright\arraybackslash}p{3.6cm}
  >{\raggedright\arraybackslash}p{3.5cm}
  >{\raggedright\arraybackslash}p{3.5cm}@{}}
\toprule
\textbf{Case} & \textbf{Source $\rightarrow$ proposition (role)} &
\textbf{Interface / test} & \textbf{Observation} &
\textbf{Relation (branch)} \\
\midrule
STM-HAR & IRB protocol $\rightarrow$ explicit package exclusions
(mechanism claim) & Package validator on 8 injected variants and the
clean package & 8/8 variants \textsc{FAIL}; 0/9 false positives &
Contradictory (variants), Consistent (clean); $q\in Q$, $\rho$ \\
STM-HAR & IRB protocol $\rightarrow$ movement/activity confidentiality
concern (acknowledged risk) & Activity inference on $G2$/$G3$ &
Adv.\ $+0.737$/$+0.785$ & No Applicable Commitment; $G=1$,
$Q=\emptyset$ \\
MIMIC & Release documentation $\rightarrow$ identifiers regenerated per
release (mechanism claim) & Released identifier fields across releases &
Operates as documented & Consistent; $q\in Q$, $\rho$ \\
MIMIC & Release documentation $\rightarrow$ 2008--2012 population
overlap (contextual fact); non-linkage stated only for identifiers & Exact $G3$
recurrence; candidate association by a credentialed recipient & 1{,}231
exact pairs; Top-1 $>97\%$ on recurrence populations only & No
Applicable Commitment; $G=1$, $Q=\emptyset$ \\
MIMIC & CareVue documentation $\rightarrow$ excludes patients with a
MIMIC-IV hospitalization (mechanism claim) & Recurrence restricted to
CareVue & $1{,}231\rightarrow0$; $7{,}886\rightarrow54$; one residual
edge unresolved & Consistent at population level; $q\in Q$, $\rho$ \\
StreetAware & Dataset publication $\rightarrow$ face-region blurring
(mechanism claim) & Face-vs.-torso sharpness & Ratio $0.264$ &
Consistent; $q\in Q$, $\rho$ \\
StreetAware & Same publication; no non-inferability statement; tracking
documented as intended (contextual fact) & Blurred-face appearance
association (R0); body$+$motion association (R1) & Corrected Top-1
$0.868$ vs.\ chance $0.500$ (non-trivial pools); gain $+0.0128$ & No
Applicable Commitment; $G=1$, $Q=\emptyset$ \\
Shepherd & Converter configuration and documentation $\rightarrow$
subject and institution metadata (conflicting evidence state) &
Deployment-evidence assessment & No authoritative replacement values &
Unresolved; evidence insufficient \\
\bottomrule
\end{tabularx}
\end{table*}
\subsubsection{STM-HAR: Candidate Interface Exposure}
\label{sec:eval_stmhar}
\paragraph{Dataset, interfaces, and adversary}
STM-HAR is a wearable-sensing research dataset. Its IRB protocol specifies a tiered
release that excludes raw media and direct identifiers, de-identified or
pseudonymous controlled access, and research use under the protocol, and
it documents confidentiality concerns about movement patterns and
activity information but no quantitative non-inferability commitment.
Participants (11) perform seven activities (walking, jogging, upstairs,
downstairs, sitting, standing, laying) with inertial sensors at five
body locations. Three candidate interfaces release progressively richer
representations of 2\,s windows: $G1$ (8 summary statistics of the
sensor's binary decision-tree output), $G2$ (8 time-domain statistics
for each accelerometer axis and its magnitude; 32 features), and $G3$
($G2$ plus the gyroscope axes and magnitude; 64 features). The adversary
trains random-forest, XGBoost, and multilayer-perceptron classifiers on
released windows, and results are reported for the best-performing
classifier. A session-preserving temporal
80/20 split assumes the adversary holds earlier labeled windows of the
same participants, so participant distinguishability is closed-set. We
report balanced accuracy (BA; chance $1/7\approx0.143$ for activity,
$1/11\approx0.091$ for participant) and activity advantage (Act.\
Adv.), BA minus $1/7$.

\paragraph{Observation}
\label{sec:eval_stmhar_residual}
\label{sec:stmhar_validation}

\begin{table}[t]
\centering\small
\caption{Activity-semantic inference and participant distinguishability
across STM-HAR candidate interfaces.}
\label{tab:stmhar_validation}
\label{tab:pid_exp1}
\begin{tabular}{lcccc}
Interface & Act. BA & Act. Adv. & Closed BA & LOAO BA \\
\midrule
$G1$ & 0.288 & $+0.145$ & 0.097 & $0.090\pm0.007$ \\
$G2$ & 0.880 & $+0.737$ & 0.816 & $0.269\pm0.150$ \\
$G3$ & 0.928 & $+0.785$ & 0.885 & $0.309\pm0.191$ \\
\bottomrule
\end{tabular}
\end{table}

In Table~\ref{tab:stmhar_validation}, $G1$ gives the weakest signal and
$G2$/$G3$ retain substantially more activity information. Because the
protocol refers to movement patterns more broadly than activity labels,
we also test \emph{participant distinguishability}, not identity
recovery, via closed-set classification (Exp.~1), cross-activity
transfer (Exp.~2), and leave-one-activity-out (LOAO) generalization
(Exp.~3; cross-activity results by activity tier in
Appendix~\ref{app:stmhar_detail}).
\label{sec:stmhar_pid}

$G1$ carries negligible participant-specific signal; $G2$/$G3$ show
distinguishability concentrated in locomotion. Transfer is strongest
within that cluster; for $G3$, jogging--locomotion transfers reach only
$1.5$--$2.3\times$ chance despite jogging being dynamic, overlapping
locomotion--static transfers ($0.9$--$2.0\times$). Held-out locomotion activities remain
distinguishable under LOAO ($\mathrm{BA}\approx0.52$--$0.54$), while
other held-out activities stay at $0.12$--$0.18$. Accelerometer
measurements account for most of the classifier-relevant signal.

\paragraph{Interpretation}
$G1$ instantiates the documented tiered-release mechanism
(\emph{Consistent}); its near-baseline exposure is a separate
observation the protocol does not bound. The acknowledged movement and
activity risk grounds the $G2$/$G3$ tests without constraining the
measured magnitude, so that exposure has \emph{No Applicable
Commitment}, as do participant distinguishability (the source does not
address participant identification) and the little reduction from
gyroscope removal ($G3\to G2$) and modality attribution (the mechanism
states no allocation across sensors). None of these establishes
open-population association, identity recovery, or deanonymization.

\subsubsection{MIMIC: Retrospective Cross-Release Audit}
\label{sec:eval_mimic}
\label{sec:eval_residual}

\paragraph{Governance evidence and interface}
No contract existed at publication, so \name\ reconstructs one
retrospectively, assuming independently prepared releases with no public
cross-release identifier mapping. The source directly documents
independent identifier regeneration, credentialed access with a
prohibition on re-identification attempts, research use including
CareVue overlap studies, the 2008--2012 population overlap, and the
CareVue mechanism that excludes individuals with a MIMIC-IV
hospitalization (Table~\ref{tab:evidence_chains}). The prohibition is a
recipient/use restriction, not an outcome guarantee that identity
recovery is technically impossible, and the overlap is a contextual fact
rather than an acknowledged privacy risk.

\paragraph{Operationalization}
\label{sec:eval_mimic_residual}
The adversary is a credentialed recipient with authorized access to both
releases but no privileged identifiers or cross-release mapping; we
measure what the released representations technically permit. Three admission-level
representations
are built identically for both releases: $G1$, demographics (age at
admission capped at 90, sex, and admission type, encoded as numeric age,
a binary sex indicator, and one-hot admission-type indicators); $G2$,
the mean, minimum, maximum, standard deviation, and count of each of the
50 laboratory concepts with the highest admission coverage (250
features); and $G3=G1\cup G2$. The evaluated populations are all
eligible admissions: 58{,}110 in MIMIC-III and 443{,}852 in MIMIC-IV. A
pair is recurrent, $\mathrm{Rec}_{\tau}(x_i^S,x_j^T)=1$, when
$O_\tau(x_i^S)=O_\tau(x_j^T)$; the target-release candidate set
$C_{\tau}(x_i^S)=\{x_j^T\mid\mathrm{Rec}_{\tau}(x_i^S,x_j^T)=1\}$
induces candidate-set reduction
\begin{equation}
\mathrm{Risk}_{\mathrm{assoc}}
=
\mathbb{E}_{i}
\left[
1-\frac{|C_{\tau}(x_i^S)|}{N_T}
\right].
\label{eq:assoc_risk}
\end{equation}

\paragraph{Procedure and controls}
Recurrence is discovered by exact equality of $G3$ across all eligible
admission pairs. Each recurring pair is checked against the underlying
raw laboratory events, including dimensions not used for discovery, and
the recurrence rate is compared with matched null controls
(uniform-random and demographic-, coverage-, count-, missingness-, and
era-matched admission pairs). Candidate association ranks all
443{,}852 MIMIC-IV admissions by Euclidean distance between standardized
representations of a source record; Top-1 is the fraction of source
records whose recurring counterpart ranks first.


\begin{table*}[t]
\centering\small
\caption{MIMIC-III/MIMIC-IV evidence for cross-release
representation recurrence, its explanation, and its operational
implications.}
\label{tab:mimic_synthesis}
\begin{tabular}{p{2.0cm} p{7.6cm} p{4.0cm}}
\toprule
\textbf{Step} & \textbf{Finding} & \textbf{Representative Evidence} \\
\midrule
Existence &
Cross-release representation recurrence is observed and is not
reproduced by the evaluated chance, covariate-matched, or
discovery-procedure controls; confirmed against raw events. &
1{,}231 exact pairs; 100\% raw-exact; 0/300 per null family;
7{,}886-edge expansion \\
Explanation &
Recurrence concentrates in the two earliest released
\texttt{anchor\_year\_group} categories, consistent with the documented
population-overlap period, and largely disappears in CareVue, an
independently curated overlap-excluding subset. &
100\% (1{,}231) / 99.35\% (6{,}655) in two earliest categories;
98.5--100\% lower recurrence rate per source admission in CareVue;
53/54 residual sparse collisions, 1 unresolved \\
Operational implication &
For records in the established recurrence populations, the recurrence
supports high-accuracy candidate retrieval under the evaluated ranking
adversary. &
Top-1 $>$97\% (recurrence populations only) \\
\bottomrule
\end{tabular}
\end{table*}

\paragraph{Interpretation}
Table~\ref{tab:mimic_synthesis} establishes representation recurrence,
population-level candidate association, and operational retrieval under
the evaluated adversary, but neither confirmed same-patient identity,
which would require external ground truth, nor a failure of the
documented de-identification procedure. The Top-1 rate is conditional
on source records from the established recurrence populations (the
1{,}231 exact pairs and the 7{,}886-edge expansion), each ranked against
all 443{,}852 MIMIC-IV admissions; it is not a population-wide linkage
rate. Identifier regeneration operates as specified (\emph{Consistent}).
The concentration in the two earliest \texttt{anchor\_year\_group}
categories (2008--2010 and 2011--2013) matches the documented overlap,
but that context grounds an explanation rather than a commitment
(\emph{No Applicable Commitment}). The CareVue collapse is
\emph{Consistent} with the separately documented exclusion mechanism at
the population level, without implying that CareVue eliminates all
cross-release association: of its 54 residual edges, 53 show the
sparse-collision pattern and one information-rich edge remains
unexplained, retained as an open observation. Recurrence and candidate
association have \emph{No Applicable Commitment}, because the
identifier-scoped protections do not govern non-identifier
representations (Appendix~\ref{app:mimic_detail}).

\subsubsection{StreetAware: Mechanism-Scoped Evaluation}
\label{sec:eval_streetaware}

\paragraph{Governance evidence and interface}
StreetAware's publication documents an IRB exemption and a face-region
anonymization mechanism (pose-guided localization, Gaussian blurring, a
secondary face detector, and conservative detection thresholds)
\cite{piadyk2023streetaware}, and identifies cross-frame tracking,
occupancy, speed, and trajectory reconstruction as intended
capabilities, so tracking alone is not evidence of nonconformance.
Phrases such as ``fully anonymized'' and ``robust video anonymization''
document the mechanism, not a facial non-inferability or
identity-nonrecovery promise. R0, within-recording association through
the appearance similarity of released blurred-face crops, is an
evaluator-defined mechanism-scoped test; R1, association through
face-excluded body appearance and motion, lies outside the mechanism's
scope and overlaps the documented tracking utility. No outcome
commitment, access/recipient condition, purpose restriction, or
acknowledged risk covers either.

\paragraph{Operationalization}
We evaluate the left camera of one sensor node in one recording of this
multi-sensor street-intersection dataset; the released face-blurred
video is the interface and the adversary's only input. A paired
face-versus-torso sharpness analysis confirms that anonymization is
present (ratio of mean Laplacian variance $0.264$, frame-clustered
bootstrap 95\% CI $[0.206,0.346]$, 118 paired detections in 40 frames).
A greedy intersection-over-union tracker builds pseudo-tracks, which
give within-recording association labels, not identity ground truth, and
are cut into temporally separated fragments; 351 tracks are retained,
and the pooled evaluation uses 5{,}481 fragment pairs from 349 of them.
Each query's candidate pool of $k\in\{2,5,10,20\}$ fragments holds its
true continuation and $k-1$ random or most-confusable (hard) negatives.
A pairwise scorer is trained and tested with track-disjoint
cross-validation; Top-1 is the fraction of queries whose true
continuation scores highest, against chance $1/k$ for a complete pool.
R0 represents only the released blurred-face region with a generic,
non-face-specialized appearance model and keeps only candidates with a
detected face, so its effective pools can be smaller than $k$ and its
reference is the mean of $1/k_{\mathrm{eff}}$. R1 excludes detected face
regions from whole-person crops before combining body appearance with
motion.

\paragraph{Face channel (R0)}
R0 is evaluated on the face-visible subset: 2{,}908 of the 5{,}481
queries have a detected face in both the query window and its true
continuation. Pools were built before face filtering, which can remove
every negative, and the historical scoring counted such positive-only
pools as Top-1 hits; we therefore report corrected Top-1 over
non-trivial pools, which retain at least one face-bearing negative. In
the primary two-candidate random-negative configuration, 1{,}067 of the
2{,}908 scored queries (36.7\%) were positive-only; on the remaining
1{,}841, corrected Top-1 is $0.868$ against an effective-pool chance of
$0.500$, a chance-adjusted excess of $0.368$ (anchor-track-clustered
bootstrap 95\% CI $[0.350,0.387]$). With two-candidate hard negatives,
922 (31.7\%) were positive-only, and corrected Top-1 is $0.789$ over
1{,}986 queries, an excess of $0.289$ ($[0.259,0.317]$). Across all
eight pool configurations the excess remains positive
($0.289$--$0.515$) and is lower for hard negatives at every $k$
(Appendix~\ref{app:streetaware_detail}). The fitted scorer orders every
query's candidates exactly as raw cosine similarity does, so R0 measures
appearance-similarity association, not a learned recognition
capability. The chance reference and intervals are descriptive, not
inferential: positives and negatives are not exchangeable by
construction, and anchor-track clustering ignores donor-fragment reuse
across queries.

This study bears on two propositions: the sharpness check is \emph{Consistent}
with the source-direct mechanism claim, while the association result
concerns a non-inferability proposition the source never states and has
\emph{No Applicable Commitment}. It establishes neither facial identity
recovery, re-identification, nor cross-camera tracking, and does not
isolate facial features from other content of the face crops.

\paragraph{Body and motion (R1)}
Combining face-excluded body appearance and motion improves association
over either alone (Top-1 $0.968$ vs.\ $0.956$/$0.939$), a composition
gain of $+0.0128$ (95\% CI $[0.0087,0.0175]$, $p<0.0001$) that grows
with temporal gap but does not persist under the smallest hard-negative
pool (gain $-0.0009$, 95\% CI $[-0.0056,0.0039]$). Excluding the face
region does not materially reduce association, so the effect is not
concentrated in the anonymized region. The target is grounded in the
documented tracking utility, noted alongside the relation rather than
treated as a requirement, and has \emph{No Applicable Commitment}. The
finding is limited to within-view pseudo-track association on one
recording and does not establish identification or cross-camera
linkage.

\subsubsection{Replication with LLM-Extracted Representations}
\label{sec:eval_spectra_rq2}

\paragraph{Design}
We recomputed nine MIMIC and StreetAware relations using each
LLM-extracted representation, r-ds (DeepSeek) and r-qw (Qwen),
separately as the governance input $C$, with
the relation procedure (Eqs.~\ref{eq:governance_relation}
and~\ref{eq:proposition_relation}), operationalizations, observations,
and frozen codebook unchanged (Table~\ref{tab:spectra_rq2}). The
interpretive inputs $\mathit{Trace}$, $\mathit{Applies}$, $\rho$, and
$\eta$ were drafted by an AI assistant in separate, non-blinded passes
per extraction (for r-qw, 451 distinct judgment inputs, or 457 rows
including six for the one record identical to an r-ds record). For
contested rules, the adjudication procedure adopted the assistant's
recommended option in all nine decisions reported here
(four on r-qw records and the five harmonization decisions below). The
assistant was a Claude model (the protocol records Opus~5.5 for the
evidence-state checks), the same family as the RQ1 sessions, so the two
steps are not independent at the model-family level, and no judgment was
independently adjudicated. An evidence state judged \emph{qualified}
counts as insufficient, yielding \emph{Unresolved}.

\begin{table}[t]
\centering
\caption{RQ2 relations for the nine operationalizations: manually reviewed comparison relations (Table~\ref{tab:evidence_chains}) and the two
LLM-extracted representations (adjudicator-confirmed common adjudication policy; the pre-harmonization adjudication yields the same relations).
NAC = No Applicable Commitment; \{C,U\} = \{Consistent, Unresolved\}.}
\label{tab:spectra_rq2}
\small
\setlength{\tabcolsep}{2.5pt}
\begin{tabularx}{\columnwidth}{@{}l>{\raggedright\arraybackslash}Xlll@{}}
\toprule
$u$ & Observation & \begin{tabular}[b]{@{}c@{}}Manually\\Reviewed\end{tabular} & DeepSeek & Qwen \\
\midrule
$u_{M1}$ & identifier-namespace overlap & Consistent & Unresolved & Unresolved \\
$u_{M2}$ & exact $G3$ recurrence & NAC & NAC & NAC \\
$u_{M4}$ & mean-only exact edges & NAC & NAC & NAC \\
$u_{M5}$ & epochs of recurring pairs & NAC & NAC & NAC \\
$u_{M6}$ & CareVue-restricted edges & Consistent$^{\dagger}$ & NAC & NAC \\
$u_{M7}$ & retrieval on recurrence populations & NAC & NAC & NAC \\
$u_{S1}$ & face/torso sharpness & Consistent & \{C,U\} & \{C,U\} \\
$u_{S2}$ & blurred-face association & NAC & NAC & NAC \\
$u_{S4}$ & body$+$motion association & NAC & NAC & NAC \\
\bottomrule
\end{tabularx}

\vspace{2pt}\footnotesize $^{\dagger}$Not comparable: the CareVue page is absent from the extraction corpus, and the
related statement is extracted as a contextual fact.
\end{table}

\paragraph{Historical and harmonized adjudication}
With judgments as first recorded, the extractions agree on all nine
relations, but the adjudications were not configuration-equivalent: r-ds
kept a pre-revision \emph{compatible} coding for two identifier
mechanisms that the codebook's coverage rule (a universally or
identifier-scoped proposition is \emph{compatible} only if the
observation covers that domain) makes \emph{cannot establish}; r-ds
excluded an anonymization outcome from the blur test under an uncodified
convention, whereas r-qw included the equivalent statement; r-ds alone
judged a procedural anonymization mechanism too unspecific to test; and
one StreetAware release-specification row received opposite
evidence-state verdicts. A single human adjudicator then confirmed a common policy
through five decisions, each the AI-recommended option: the written
coverage rule applies to both extractions (H1); applicability follows
the written test, which includes the anonymization outcome in the blur
test (H2); a term defined elsewhere in the source is a matter of
applicability rather than evidence support (H3); the anonymization
sensitivities cover all equivalent records (H4); and a mechanism whose
operative term the source defines elsewhere is testable by the measured
property, so the procedural mechanism applies in both extractions (H5).
This changed five r-ds judgments (two under H1, one each under H2, H3,
and H5), no r-qw judgment, and no relation; the historical rows keep
their never individually approved status, with the five changes recorded
separately as adjudicator-confirmed. Under the common policy, the extractions agree on 9/9
operationalizations and each matches PACT's manually reviewed relation on 6 of
the 8 comparable ones. This is agreement on final relations from one
AI-assisted judgment process, and seven of the nine are \emph{No Applicable
Commitment}, which requires only that neither extraction place a
supported proposition in $Q$.

\paragraph{Routes to the two differences}
Both differences are more conservative than the manually reviewed
\emph{Consistent}. For $u_{S1}$, both extractions' propositions are
supported, but ``fully anonymized'' is universally quantified under the
coverage rule while one recording has unexplained frames, giving
\{Consistent, Unresolved\}. For $u_{M1}$, both reach \emph{Unresolved}
through different branches of Eq.~\ref{eq:governance_relation}, because
the same non-linkage sentence, with the same text and span, carries
different roles. r-ds encodes it as an applicable outcome commitment
whose extracted proposition drops the source's causal and concessive
qualifiers; its evidence is judged \emph{qualified}, so the procedure
stops at the evidence-sufficiency branch and masks two identifier
mechanisms that are \emph{cannot establish} under the coverage rule
(lifting that check still yields \emph{Unresolved}). r-qw encodes it as
a contextual fact, which the procedure's role gate (contextual facts and
acknowledged risks never apply) keeps out of $Q$, so the relation
aggregates the two mechanisms, \emph{cannot establish} because they are
scoped to subject\_id or all patient identifiers while the observation
measures only hadm\_id. Extraction thus fixes the available semantics
and roles, roles fix applicability, and adjudication assigns evidence
states; the coverage rule supplies $\rho$ inputs and is not part of the
procedure. Identical final relations do not require identical
intermediate paths. The comparison with the manually reviewed baseline is also
confounded by records, corpus, and codebook revision, so neither difference can be
apportioned among these factors, and cross-extraction agreement
establishes neither agreement with the manually reviewed baseline, semantic equivalence,
nor correctness.

\paragraph{Proposition-level differences}
Same-role matching with overlapping evidence pairs 405 r-qw rows with
r-ds records (426 combinations); role differences lie outside these
counts, and 52 of 457 r-qw and 58 of 468 r-ds rows are unmatched, only
one of them (the r-ds non-linkage outcome for $u_{M1}$) with
$\mathit{Applies}=1$. Harmonization reduced applicability and
compatibility differences from 4 to 0 by changing only r-ds rows; as the
r-qw identifier-mechanism values had been set earlier by adjudicator decisions
under the same rule, this holds by construction and is not independent
evidence of convergence. Evidence-state differences among the 21
combinations assessed in both extractions fell from 3 to 0 (23 others
are not comparable). Seventy-two combinations (67 r-qw rows) still
differ in $\mathit{Trace}$ only, 60 with r-ds at 0 (48 of them
never-adjudicated r-ds default rows without an $\mathit{Applies}$ value)
and 12 the reverse. None has $\mathit{Applies}=1$ on either side and
other $\mathit{Trace}=1$ rows fix $G$, so they cannot change a relation
as recorded, but they remain unreconciled.


\paragraph{Adjudication sensitivity}
\label{sec:eval_spectra_sensitivity}
Nine targeted configurations, not a factorial design, each apply one
change to the judgment inputs of \emph{both} extractions and re-run the
unmodified engine (Table~\ref{tab:spectra_sensitivity}). Four were
defined before execution (``Pre''): under the non-identifiability and
the explicitly hypothetical strong-guarantee readings, anonymization
outcome records apply to $u_{S2}$ and $u_{S4}$ as \emph{cannot
establish} or to $u_{S2}$ as \emph{conflicts}, and the CareVue
counterfactual makes that record a mechanism \emph{compatible} with
$u_{M6}$. An adjudicator-approved prospective amendment widened both readings
to all equivalent outcome records, the r-qw CareVue counterpart was
added later, and its pre-defined \emph{cannot establish} variant was not
re-run; the other five were added after the audit as alternatives to the
harmonization decisions. Eight configurations keep 9/9, some moving a
relation in both extractions together. The exception, an exploratory
reading that contradicts the written coverage rule by treating the
partial identifier measurement as compatible, makes $u_{M1}$
\emph{Consistent} for r-qw but leaves it \emph{Unresolved} for r-ds
(8/9).

\begin{table}[t]
\centering
\caption{Symmetric adjudication sensitivity: cross-extractor agreement between the DeepSeek and Qwen representations on the nine
operationalizations. ``Pre'' marks configurations defined before execution.}
\label{tab:spectra_sensitivity}
\small
\setlength{\tabcolsep}{3pt}
\begin{tabularx}{\columnwidth}{@{}>{\raggedright\arraybackslash}X l c >{\raggedright\arraybackslash}p{2.3cm}@{}}
\toprule
Change (both extractions) & Varied & Agree & Relation changed \\
\midrule
Qualified evidence as supported (Pre) & $\eta$ & 9/9 & none \\
Term defined elsewhere $\Rightarrow$ qualified & $\eta$ & 9/9 & $u_{S1}$ \emph{Unresolved} \\
Outcome records excluded from blur test & Applies & 9/9 & none \\
Procedural mechanism excluded from blur test & Applies & 9/9 & none \\
Outcome records excluded \emph{and} term defined elsewhere $\Rightarrow$ qualified & both & 9/9 & $u_{S1}$ \emph{Unresolved} \\
Non-identifiability reading (Pre) & rule & 9/9 & $u_{S2},u_{S4}$ \emph{Unresolved} \\
Strong-guarantee reading (Pre) & rule & 9/9 & $u_{S2}$ \emph{Contradictory} \\
CareVue record as mechanism (Pre) & role & 9/9 & $u_{M6}$ \emph{Consistent} \\
Partial identifier measurement compatible & $\rho$ & \textbf{8/9} & $u_{M1}$ differs \\
\bottomrule
\end{tabularx}
\end{table}

\paragraph{Tracing Divergent Evaluation Paths}
Under this exploratory reading, the two extractions produce
different relations for $u_{M1}$ through distinct evaluation
paths: the r-ds relation is decided at the evidence-sufficiency
step, whereas the r-qw relation aggregates two mechanisms
that the reading makes compatible.
Single-record counterfactuals, run as diagnostics on copies
of the inputs rather than as alternative results, indicate
that the disagreement arises from an extraction-induced role
difference interacting with role-dependent applicability,
an adjudicated evidence-state verdict, and the procedure's
precedence of evidence sufficiency over compatibility,
not from extraction alone.
The role difference alone does not account for the
disagreement in every counterfactual direction.

\subsection{RQ3: Prospective Validation}
\label{sec:eval_rq3}
\label{sec:eval_prospective}

We evaluate prospective use of source-supported propositions at the
STM-HAR release boundary and the BBQS upload boundary.

\paragraph{STM-HAR release validation}

Before any release, we derived nine package-level rules from the
STM-HAR governance record, checking source-supported constraints on
structure, schemas, prohibited identifiers and filenames, excluded
media, and metadata. On the clean candidate package, 8/9 rules PASS and
1/9 produces a WARNING because the source specifies no corresponding
requirement; none FAIL.
All 8/8 controlled variants containing content excluded by applicable
source-supported propositions are detected, with 0/9 false positives on
the clean package. These variants are the positive-control
\emph{Contradictory} cases because they directly violate explicit
applicable exclusions; successful inferential attacks alone do not.

\paragraph{BBQS upload validation}
Our BBQS prototype binds an asserted authority reference and target to
the package inventory before an upload proceeds, using synthetic
authority values, and rejects missing or inconsistent bindings under the
evaluated tests. The check executes at the client and may be bypassed by
a modified client or alternate upload path, so it establishes
prospective interface validation, not archive-side or end-to-end
enforcement.
\subsection{RQ4: Operational Interfaces}
\label{sec:eval_system}
\label{sec:eval_rq4}

At operational configuration, upload, and conversion interfaces,
reported separately from the observations above, we distinguish
\emph{technical capability}, \emph{observed deployment},
\emph{governance authorization}, and \emph{enforcement}, never promoting
evidence for one into evidence for another (Table~\ref{tab:system_eval}).

\paragraph{CAPTURE configuration sharing}
\label{sec:eval_capture}
CAPTURE is a research sensing platform with a public repository
\cite{captureGithub}. The evaluated administrative component
configures sensing deployments. Its portable configuration boundary could also
carry HTTP-push state, coupling reusable sensing settings to
deployment-specific delivery information. A candidate revision omits HTTP-push details from
shared configurations so that importing preserves the receiving device's
existing state. We do not claim that it has been merged, adopted, or
deployed; it is a candidate refinement, not fail-closed protection or
deployment-wide enforcement.

\paragraph{BBQS upload and conversion}
BBQS is a multi-site neuroscience consortium whose browser uploader
transfers study datasets to a shared archive (its signed preflight,
Section~\ref{sec:eval_rq3}, remains a client-side check) and whose
ingest pipelines convert raw recordings into standardized formats. The
Shepherd pipeline, which converts multicamera behavioral recordings into
NWB representations \cite{bbqsShepherdGithub}, is a complementary
negative case: the pinned converter source has a subject species value
that conflicts with the project's own documentation and an institution
value stored under a configuration key the converter does not read; no
authoritative replacement values were available, and we observed no
real-session conversion. This evidence cannot determine whether any
deployment or study claim holds, so the stronger claim remains
\emph{Unresolved}.

\begin{table}[t]
\centering
\small
\caption{System-facing evaluation of CAPTURE~\cite{captureGithub},
BBQS, and Shepherd. \name\ keeps evidence of software capability
separate from deployment, authorization, and enforcement.}
\label{tab:system_eval}
\begin{tabular}{p{1.45cm}p{1.8cm}p{2.05cm}p{1.65cm}}
\toprule
\textbf{Interface} &
\textbf{Observed evidence} &
\textbf{PACT outcome} &
\textbf{Claim boundary} \\
\midrule

CAPTURE config. share/import &
Non-public implementation evaluation + candidate revision &
Portable and deployment-local delivery state can be separated &
Not an independent assessment; candidate, not adopted or fail-closed \\

BBQS upload &
Prototype + bounded validation &
Signed preflight binds authority, target, and inventory &
Client-side; not end-to-end enforcement \\

Shepherd conversion &
Pinned source + project documentation &
Unresolved: conflicting metadata &
No observed real-session conversion \\

\bottomrule
\end{tabular}
\end{table}

\section{Discussion}
\label{sec:discussion}

\paragraph{Calibrating claims}
Attack success, governance contradiction, and compliance need different
evidence: an attack shows what an interface reveals under stated
assumptions, a contradiction also requires a supported proposition whose
scope covers the measured property, and compliance involves legal and
procedural determinations that \name\ does not make. In StreetAware,
correcting the face-channel evaluation lowered the measured association
but left the relation unchanged, because no source commitment covers
that property. \emph{Consistent} is not a privacy score, and \emph{No
Applicable Commitment} is not evidence that no risk exists: it marks
exposure the reviewed governance does not address, often the most
actionable finding for a custodian. Authors who intend an outcome
commitment, rather than a mechanism such as identifier regeneration or
face blurring, can state it with its population, recipient, release, and
temporal scope so that it can be tested and, where appropriate,
contradicted; recorded provenance and evidence state make differing
interpretations traceable to a source span and role. At operational
interfaces, a check matters only where it binds: a client-side
preflight is not archive-side enforcement.

\paragraph{Automatically constructed representations as inputs}
PACT's contribution remains the evidence-bounded interpretation above;
automated extraction concerns how the evidence is constructed. The
replications
support a limited claim: \emph{in these two retrospective cases, under
one reconstruction model and an adjudicator-confirmed adjudication policy,
LLM-extracted representations served as alternative governance inputs
within the PACT workflow while largely preserving reconstruction
decisions and downstream relations}, not that automated extraction can
replace PACT's evidence construction. Retaining the manually reviewed records
and the 9/9 cross-extractor agreement are
not contradictory: one tests replacement
under a strict unanimity rule, the other whether two extractions lead
the fixed procedure to the same relations, and the two differences from
the manually reviewed relations are more conservative, not equivalent, outcomes. In these
nine operationalizations, extraction variation acted mainly through
which propositions exist and which roles they carry, changing the route
to a relation rather than the relation; stable final relations can
therefore conceal different intermediate reasoning, so we report routes
alongside relations and keep extracted roles visible rather than
harmonizing them. The results establish neither accuracy (no ground-truth
answers exist, so higher consistency does not imply higher correctness),
superiority of either input, independent validation, nor
generalization beyond the two
retrospective cases.

\paragraph{Limitations and future work}
\name\ yields bounded interpretations, not exhaustive privacy
guarantees, and each case is scoped accordingly. MIMIC lacks cross-release identity
ground truth, so cross-release association establishes neither
same-patient linkage nor re-identification. STM-HAR is bounded to its
cohort, activities, and closed-set distinguishability, and is not an
exhaustive privacy evaluation. StreetAware covers one recording,
session, and camera with pseudo-track labels that may contain tracking
errors; its results depend on negative sampling, and its face-channel
result, limited to the face-visible subset, may reflect non-facial crop
content and positive--negative differences, with confidence intervals
that do not account for shared-donor dependence (Section~\ref{sec:eval_streetaware}).
None of the cases establishes identity recovery or cross-camera
tracking, and PACT cannot infer absent obligations or establish legal
compliance or privacy safety.

The representation analyses are secondary to the relation procedure.
The six-run blinded replication evaluates only procedural consistency
(Section~\ref{sec:eval_rq1}), not independent correctness, exhaustive
identical extraction, or generalization to arbitrary governance prose.
The manually reviewed records, initially constructed with LLM
coding-agent assistance, are not independent human-authored ground
truth. The LLM-extraction comparisons do not establish generalizability
across models or extraction conditions; agreement neither demonstrates
correctness nor justifies replacing manually reviewed records. Future work,
none evaluated here, includes
additional datasets and governance genres, such as data-use agreements
and consent forms; human inter-rater studies of role and scope
assignment and validation of automated representations against blinded
human adjudication; and server-side enforcement points.

\paragraph{Limitations of the representation comparison}
Beyond the caveats that apply to the results in
Sections~\ref{sec:eval_rq1} and~\ref{sec:eval_spectra_rq2} (one model,
unequal inputs, a different corpus, post hoc AI-assisted adjudication
confirmed by a single adjudicator, five
sensitivity configurations added after the audit, unreconciled
$\mathit{Trace}$ differences, and a confounded comparison with the
manually reviewed baseline), the RQ1 sessions differ from the blinded
ChatGPT/Gemini sessions in interface (Claude Code CLI, which adds a
short system prompt and environment preamble) and execution context; the
per-scenario evidence-state overrides were recovered by a byte-identical
re-execution rather than persisted at run time; the r-qw results
(the Qwen condition and the r-qw replays) were verified only against
author-held files; the causal account of the 8/9 result rests on
single-record diagnostic counterfactuals; and the earliest extraction
sequence cannot be replayed because its agent-session transcripts were
not retained. The replications cover only
MIMIC and StreetAware, and the counts carry no significance test and do not
generalize to other governance genres or extraction pipelines.

\section{Conclusion}
\label{sec:conclusion}

Privacy exposure alone does not establish a governance violation.
\name\ is a provenance-aware framework that connects governance evidence
to executable data-interface evaluations and interprets observations
only against applicable, source-supported propositions. Across wearable
sensing, clinical-data releases, and anonymized video, our evaluations
validate explicit release exclusions where specified and show that
substantial inferential exposure can coexist with documented mechanisms
operating as described; at operational interfaces, \name\ separates
implementation capability from deployment, authorization, and enforcement.

LLM-extracted representations yielded substantial agreement in
downstream interpretations despite differing evidence construction and
adjudication, supporting representation-sensitivity analysis without
establishing independent validation or justifying replacement of
manually reviewed records. Together, these results show the need to
distinguish what an empirical test reveals from what a documented
commitment requires. By preserving each proposition's provenance and
scope and separating documented commitments from evaluator assumptions,
\name\ provides a structured basis for relating observations to
applicable commitments, enabling transparent, appropriately bounded
privacy and security assessments.



\bibliographystyle{IEEEtran}
\bibliography{\pactsourcepath ref}


\appendices

%
\providecommand{\tablestyle}{\small}
\providecommand{\thead}[1]{\textbf{#1}}
\providecommand{\rowgap}{\addlinespace[2pt]}
\providecommand{\pactsourcepath}{}
\makeatletter
\@ifundefined{NC@find@L}{\newcolumntype{L}{>{\raggedright\arraybackslash}X}}{}
\@ifundefined{NC@find@P}{\newcolumntype{P}[1]{>{\raggedright\arraybackslash}p{#1}}}{}
\makeatother
\newenvironment{appendixpage}{\begin{table*}[p]\normalsize}{\end{table*}}
\newenvironment{appendixtable}{%
  \par\addvspace{\intextsep}\begin{minipage}{\linewidth}%
}{\end{minipage}\par\addvspace{\intextsep}}
\makeatletter
\newenvironment{appendixfigure}{%
  \par\addvspace{\intextsep}\begin{minipage}{\linewidth}\def\@captype{figure}%
}{\end{minipage}\par\addvspace{\intextsep}}
\makeatother

\begin{appendixpage}
\section{STM-HAR Supporting Results}
\label{app:stmhar_detail}
\label{app:additional_results}


\begin{minipage}[t]{\columnwidth}\vspace{0pt}
\paragraph{Contract provenance}
Table~\ref{tab:stmhar_contract_stats} summarizes the provenance of the
23 fields extracted into the STM-HAR interface contract.

\paragraph{Cross-activity distinguishability}
Table~\ref{tab:pid_exp2} provides the detailed cross-activity results
for the $G3$ interface. Distinguishability remains strongest across
locomotion activities, decreases for jogging, and approaches the random
baseline for several static-posture transfers.

The modality-attribution analysis assigns accelerometer shares of
$1.000$ for $G2$ and $0.666$ for $G3$; $G1$ has no corresponding
accelerometer-share value.
\end{minipage}\hfill
\begin{minipage}[t]{\columnwidth}\vspace{0pt}
\paragraph{Evidence-bounded interpretation}
Table~\ref{tab:stmhar_classification} relates the STM-HAR observations
to their applicable governance propositions and resulting relations.
\end{minipage}

\begin{minipage}[t]{\columnwidth}\vspace{0pt}
\begin{appendixtable}
\centering\tablestyle
\caption{Provenance summary for the STM-HAR interface contract
(23 extracted fields).}
\label{tab:stmhar_contract_stats}
\begin{tabularx}{\columnwidth}{@{}Lr@{}}
\toprule
\thead{Statistic} & \thead{Count} \\
\midrule
Total extracted fields & 23 \\
\quad Source-direct (directly supported by protocol text) & 14 \\
\quad Interpretation-derived & 9 \\ \rowgap
Unresolved fields (not specified by the governance source) & 5 \\
\bottomrule
\end{tabularx}
\end{appendixtable}
\end{minipage}\hfill
\begin{minipage}[t]{\columnwidth}\vspace{0pt}
\begin{appendixtable}
\centering\tablestyle
\caption{Experiment 2: cross-activity participant distinguishability
for $G3$ (transfers between different activities), grouped by tier:
within locomotion, jogging--locomotion, and locomotion--static posture.
Random baseline $=0.091$.}
\label{tab:pid_exp2}
\begin{tabularx}{\columnwidth}{@{}LP{2.0cm}P{1.45cm}@{}}
\toprule
\thead{Activity tier} & \thead{Cross-activity BA} & \thead{vs.\ baseline} \\
\midrule
Locomotion (Downstairs, Upstairs, Walking) &
0.38--0.60 (pairwise) & 4.2--6.6$\times$ \\ \rowgap
Jogging &
0.13--0.21 & 1.5--2.3$\times$ \\ \rowgap
Static posture (Sitting, Standing, Laying) &
0.08--0.18 & 0.9--2.0$\times$ \\
\bottomrule
\end{tabularx}
\end{appendixtable}
\end{minipage}

\centering\tablestyle
\par\addvspace{\intextsep}
\caption{Evidence-bounded interpretation of the STM-HAR evaluation
observations.}
\label{tab:stmhar_classification}
\begin{tabularx}{\textwidth}{@{}P{3.9cm}P{3.4cm}P{2.1cm}L@{}}
\toprule
\thead{Observation} & \thead{Evidence basis} & \thead{Relation} & \thead{Interpretation} \\
\midrule
$G1$ tiered-release mechanism application and near-baseline activity
semantics &
Tiered-release mechanism claim (source-direct) &
Consistent (mechanism); No Applicable Commitment (magnitude) &
The mechanism is applied as documented; the protocol states no
quantitative bound on the separately measured exposure. \\ \rowgap
$G2$/$G3$ retain substantial activity-semantic information
(Adv.\ $+0.737$/$+0.785$) &
Acknowledged risk: movement patterns/activity information
(source-direct) &
No Applicable Commitment &
The acknowledged risk grounds the empirical question but does not
constrain the measured magnitude; no applicable outcome commitment is
stated. \\ \rowgap
Closed-set, cross-activity, and leave-one-activity-out participant
distinguishability at $G2$/$G3$ &
Same acknowledged risk; evaluator-defined participant-level framing &
No Applicable Commitment &
Operationalizes a participant-level interpretation of the acknowledged
movement-pattern concern; the source addresses activity/movement
information generally, not participant identification specifically. \\ \rowgap
$G3\rightarrow G2$ gyroscope-removal minimization produces little
reduction in measured exposure &
Tiered-release mechanism claim; no per-modality allocation stated &
No Applicable Commitment &
The mechanism claim describes tiered minimization generally but
commits to no quantitative reduction from any single modality's
removal. \\ \rowgap
Accelerometer/gyroscope attribution &
No modality-level risk allocation stated &
No Applicable Commitment &
Informative for interface design, but the governance source provides no
modality-level allocation against which the result could be judged. \\
\midrule
8/8 injected fault variants correctly flagged \textsc{Fail}; 0/9 false
positives on the clean package &
Explicit prohibited-content exclusions (mechanism claim, source-direct) &
Contradictory (injected variants); Consistent (clean package) &
The manuscript's positive-control demonstration: each fault variant
directly contradicts the contract's explicit exclusions, and the clean
package correctly shows no contradiction. \\
\bottomrule
\end{tabularx}
\end{appendixpage}

\begin{appendixpage}
\section{MIMIC Supporting Evidence}
\label{app:mimic_detail}

Table~\ref{tab:rq1_full} traces representation recurrence through raw-event
validation, matched-null controls, and the independently defined Control-K
expansion (existence), documented cross-release provenance (explanation),
and operational analyses (operational implication), the three steps of
Table~\ref{tab:mimic_synthesis}. Table~\ref{tab:mimic_classification} relates these observations
to applicable governance propositions and resulting relations.
\par\addvspace{\intextsep}

\centering\tablestyle
\caption{MIMIC-III/MIMIC-IV evidence chain: validation of cross-release
representation recurrence (existence), its consistency with documented
cross-release provenance (explanation), and its operational implications.}
\label{tab:rq1_full}
\label{tab:rq1_detail}
\label{tab:rq2_full}
\label{tab:rq2_detail}
\label{tab:rq3_full}
\begin{tabularx}{\textwidth}{@{}P{3.3cm}LL@{}}
\toprule
\thead{Analysis} & \thead{Finding} & \thead{Representative evidence and supporting metrics} \\
\midrule
\multicolumn{3}{@{}l}{\textit{Existence: validation of cross-release representation recurrence}} \\
Establish high-dimensional representation recurrence &
A substantial population of admission pairs matches exactly on the
evaluated released aggregate. &
1{,}231 exact-match pairs; --- \\ \rowgap
Validate recurrence against source measurements &
The corresponding pairs also match in the underlying raw clinical
events, including dimensions not used by the discovery procedure. &
100\% raw-exact; Raw-event validation: evidence score mean $0.996$ (minimum $0.896$)
across all 1{,}231/1{,}231 pairs \\ \rowgap
Evaluate recurrence under matched null controls &
None of the chance or covariate-matched controls reproduces the
observed recurrence. &
0/300 per family ($\sim$6M nulls) across 10 matched-null families; because
no control event occurs, the risk ratio ($\approx602$, 95\% CI 38--9{,}599)
and odds ratio ($\approx1.48$M, 95\% CI $2.9\times10^{4}$--$7.5\times10^{7}$)
relative to candidates depend on a Haldane--Anscombe $+0.5$ correction \\ \rowgap
Evaluate recurrence beyond exact-$G3$ pairs &
An independently defined coarser criterion identifies a larger
population exhibiting comparable raw-event fidelity. &
7{,}886 edges (6{,}655 new); Control-K expansion: 99.86\% raw-exact and
97.7\% clean one-to-one graph structure among the 6{,}655 new edges \\ \rowgap
\midrule
\multicolumn{3}{@{}l}{\textit{Explanation: consistency with documented cross-release provenance}} \\
Temporal provenance &
Recurring pairs are concentrated almost entirely in the earliest
released \texttt{anchor\_year\_group} categories, consistent with the
documented 2008--2012 overlap period. &
100\% (1{,}231) in the two earliest categories; \texttt{anchor\_year\_group}
is a released three-year categorical field whose categories extend
beyond 2012; 99.35\% of the 6{,}655 newly identified expansion edges fall
within its two earliest categories \\ \rowgap
Overlap-restricted counterfactual &
Restricting the analysis to CareVue sharply reduces the observed
recurrence, consistent with the overlap-based explanation. &
98.5--100\% collapse; Exact-$G3$ discovery: $1{,}231\rightarrow0$; mean-only discovery:
$7{,}886\rightarrow54$; 52/54 (96.3\%) of the remaining cases are
raw-exact \\ \rowgap
Residual recurrence &
Nearly all remaining CareVue cases exhibit the sparse-collision pattern
identified in the existence analysis; one information-rich case remains
unresolved. &
53/54 cases exhibit a sparse-hub collision pattern and 1/54 remains
unresolved; 43/6{,}655 (0.65\%) of the newly identified expansion edges
fall in later \texttt{anchor\_year\_group} categories (2014--2022) \\ \rowgap
\midrule
\multicolumn{3}{@{}l}{\textit{Operational implication of validated recurrence}} \\
Candidate retrieval &
For source records in the established recurrence populations,
validated recurrence supports operational retrieval, with Top-1
accuracy exceeding 97\%, including held-out concepts; this is not a
population-wide linkage rate. &
Top-1 $>$97\% (recurrence populations only) \\ \rowgap
\bottomrule
\end{tabularx}
\par\addvspace{\intextsep}

\caption{Evidence-bounded interpretation of the MIMIC-III/MIMIC-IV
evaluation observations.}
\label{tab:mimic_classification}
\begin{tabularx}{\textwidth}{@{}P{3.9cm}P{3.4cm}P{2.1cm}L@{}}
\toprule
\thead{Observation} & \thead{Evidence basis} & \thead{Relation} & \thead{Interpretation} \\
\midrule
Identifier independence
(\texttt{subject\_id}/\texttt{hadm\_id} regeneration) &
Mechanism claim (source-direct) &
Consistent &
Independent identifier regeneration is documented and observed to
operate as specified. \\ \rowgap
Representation-level recurrence
(1{,}231 pairs; 7{,}886-edge expansion) &
Non-linkage statement scoped to identifiers only; evaluator-defined target &
No Applicable Commitment &
The documented non-linkage statement concerns linkage using an
identifier such as \texttt{subject\_id}; no source statement requires
or prohibits recurrence of non-identifier representations across
independently prepared releases. \\ \rowgap
Recurrence concentrated in the documented overlap window
(provenance/CareVue explanation) &
Contextual fact (source-direct: 2008--2012 population overlap) &
No Applicable Commitment &
The overlap grounds the empirical explanation but is not a commitment
that can be contradicted or satisfied; the source does not characterize
it as an acknowledged privacy risk. \\ \rowgap
CareVue-restricted population
($1{,}231\rightarrow0$; $7{,}886\rightarrow54$) &
Mechanism claim: exclusion of individuals with a MIMIC-IV
hospitalization; evaluator-defined counterfactual &
Consistent &
Restricting the analysis to CareVue substantially attenuates the
observed recurrence, consistent with the documented patient-exclusion
mechanism. This counterfactual is not itself a source-stated outcome
commitment, purpose restriction, or acknowledged privacy risk. The
relation concerns the exclusion mechanism, not elimination of all
cross-release association; one information-rich residual edge remains
unresolved. \\ \rowgap
Candidate association
(Top-1 $>$97\% on recurrence populations) &
Non-linkage statement scoped to identifiers only; evaluator-defined target &
No Applicable Commitment &
Quantifies the operational consequence of the documented overlap and
representation recurrence under the evaluated adversarial model,
without an applicable commitment the observation could contradict. \\
\bottomrule
\end{tabularx}
\end{appendixpage}

\begin{appendixpage}
\section{StreetAware Supporting Results}
\label{app:streetaware_detail}

\begin{minipage}[t]{\columnwidth}\vspace{0pt}
\paragraph{Governance evidence}
StreetAware documents an IRB exemption, an explicit privacy motivation,
face-scoped Gaussian blurring with secondary face detection, and intended
capabilities: tracking, occupancy and speed estimation, trajectory
reconstruction, and consistent detection across frames and views. The
source states no outcome commitment, recipient obligation, purpose
restriction, or acknowledged residual risk relevant to the evaluated
association task. Accordingly, R0's non-inferability question is an
evaluator-defined, mechanism-scoped test motivated by the documented
privacy motivation and anonymization mechanism, not a reconstructed
source commitment; \name\ records no outcome-commitment field for
StreetAware rather than inferring one from that motivation alone.

\paragraph{Experimental protocol}
Pseudo-track construction, temporal-fragment generation, and
positive/random-negative/hard-negative candidate construction use a
common greedy intersection-over-union tracker and fixed quality filters.
The appearance representation is varied among whole-person,
face-excluded body, and face-only crops while the remaining protocol is
held constant. Appearance features are extracted using an OSNet~x0.25
backbone. Evaluation uses grouped track-disjoint cross-validation,
label-permutation and time-offset controls, track-quality stratification,
and grouped-bootstrap confidence intervals with 2,000 resamples.

\paragraph{Appearance--motion composition}
Table~\ref{tab:streetaware_hardneg} reports the R1 (body and motion) composition gain
across candidate-set sizes and random versus hard negatives;
Figure~\ref{fig:streetaware_gap} shows the same comparison across
temporal gaps.

\begin{appendixtable}
\centering\tablestyle
\caption{StreetAware R1: Top-1 composition gain from body appearance
and motion across candidate-set sizes and negative types, pooled over
temporal gaps.}
\label{tab:streetaware_hardneg}
\begin{tabular*}{\columnwidth}{@{\extracolsep{\fill}}lrrc@{}}
\toprule
\thead{Negatives} & \thead{Pool} & \thead{Gain} & \thead{95\% CI} \\
\midrule
Hard   &  2 & $-0.0009$ & $[-0.0056,\ 0.0039]$ \\
       &  5 & $+0.0173$ & $[\phantom{-}0.0097,\ 0.0254]$ \\
       & 10 & $+0.0449$ & $[\phantom{-}0.0343,\ 0.0500]$ \\
       & 20 & $+0.0431$ & $[\phantom{-}0.0355,\ 0.0512]$ \\ \rowgap
Random &  2 & $+0.0128$ & $[\phantom{-}0.0087,\ 0.0175]$ \\
       &  5 & $+0.0283$ & $[\phantom{-}0.0224,\ 0.0348]$ \\
       & 10 & $+0.0283$ & $[\phantom{-}0.0212,\ 0.0360]$ \\
       & 20 & $+0.0232$ & $[\phantom{-}0.0151,\ 0.0320]$ \\
\bottomrule
\end{tabular*}
\end{appendixtable}

Only the two-candidate hard-negative interval includes zero. Gains remain positive under
track-quality stratification ($+0.0080$ high-density, $+0.0204$
medium-density). The time-offset control reduces the mean pairwise score
from $0.815$ to $0.525$ ($p<0.0001$), indicating dependence on temporally
corresponding motion information.
\end{minipage}\hfill
\begin{minipage}[t]{\columnwidth}\vspace{0pt}
\begin{appendixfigure}
\centering
\includegraphics[width=\columnwidth]{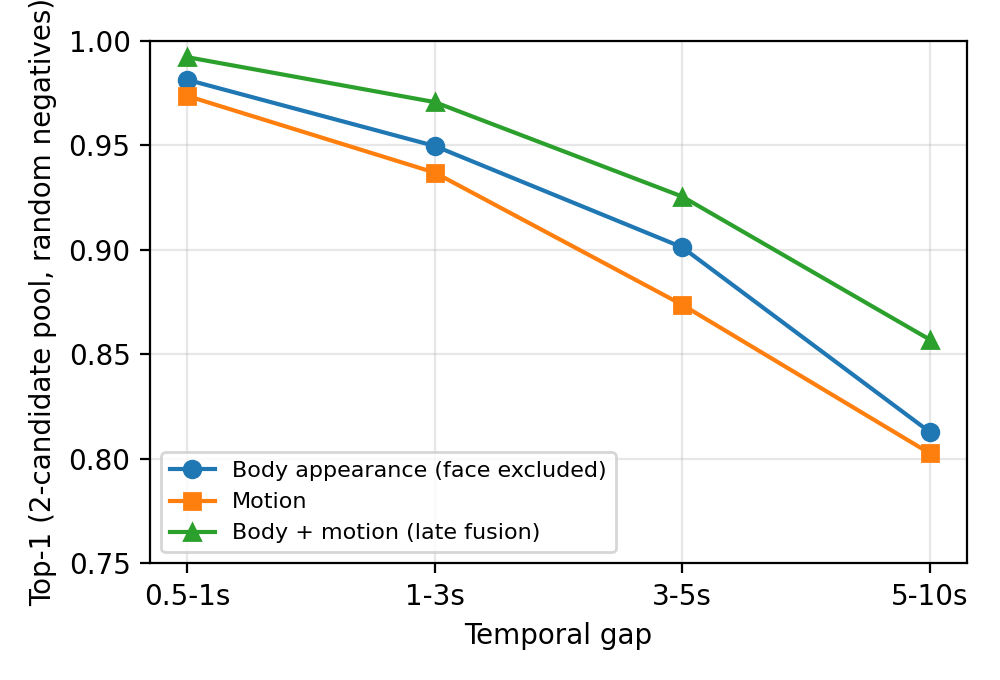}
\caption{StreetAware fragment-association performance for body
appearance, motion, and their late-fusion composition across temporal
gaps.}
\label{fig:streetaware_gap}
\end{appendixfigure}

\paragraph{Face-channel (R0) analysis}
Table~\ref{tab:streetaware_facecheck} reports the corrected R0
evaluation. Pools were built before face filtering, so the effective
pool $k_{\mathrm{eff}}$ can be smaller than $k$, and positive-only pools
($k_{\mathrm{eff}}=1$) are excluded from corrected Top-1. Configurations with fewer
than four groups are excluded, not imputed. Excluding the detected face
region from whole-person appearance changes Top-1 by only 0.02--0.9
percentage points across the eight configurations.

\begin{appendixtable}
\centering\tablestyle
\caption{StreetAware R0 on the face-visible subset (2{,}908 queries per
configuration). Historical Top-1 counts positive-only pools as hits;
corrected Top-1, chance (mean $1/k_{\mathrm{eff}}$), and excess use the
$n$ non-trivial pools. Intervals: 95\% CIs for the excess from 10{,}000
anchor-track bootstrap resamples.}
\label{tab:streetaware_facecheck}
\begin{tabular*}{\columnwidth}{@{\extracolsep{\fill}}lrrrrrc@{}}
\toprule
\thead{Neg.} & $\boldsymbol{k}$ & $\boldsymbol{n}$ & \thead{Hist.} &
\thead{Corr.} & \thead{Chance} & \thead{Excess [95\% CI]} \\
\midrule
Random &  2 & 1{,}841 & 0.916 & 0.868 & 0.500 & 0.368 [0.350, 0.387] \\
       &  5 & 2{,}853 & 0.782 & 0.778 & 0.300 & 0.478 [0.453, 0.505] \\
       & 10 & 2{,}908 & 0.671 & 0.671 & 0.156 & 0.515 [0.481, 0.552] \\
       & 20 & 2{,}908 & 0.567 & 0.567 & 0.078 & 0.489 [0.451, 0.533] \\ \rowgap
Hard   &  2 & 1{,}986 & 0.856 & 0.789 & 0.500 & 0.289 [0.259, 0.317] \\
       &  5 & 2{,}790 & 0.724 & 0.713 & 0.295 & 0.418 [0.382, 0.452] \\
       & 10 & 2{,}897 & 0.621 & 0.619 & 0.164 & 0.455 [0.417, 0.495] \\
       & 20 & 2{,}908 & 0.528 & 0.528 & 0.083 & 0.446 [0.402, 0.495] \\
\bottomrule
\end{tabular*}
\end{appendixtable}

\paragraph{Scoring and reference}
The fitted scorer orders every non-trivial query's candidates exactly as
raw cosine similarity does, so R0 measures appearance-similarity
association. An earlier label-permutation null (mean $0.685$ in the
primary configuration) is not a no-information reference: permuted fits
still rank candidates by similarity, and positive-only pools count as
hits in both arms. It is therefore excluded from the reported results.
\end{minipage}
\end{appendixpage}

\clearpage
\section{Ethical Considerations}
\label{app:ethics}

\paragraph{Research Scope and Human-Subjects Data}
The empirical privacy experiments on STM-HAR, MIMIC-III/MIMIC-IV, and
StreetAware are conducted offline on previously collected, candidate,
or released data. These experiments do not alter clinical records,
contact dataset participants, or make automated decisions about
individuals. STM-HAR is governed by an active institutional IRB
protocol; MIMIC-III/MIMIC-IV are de-identified clinical datasets
accessed under applicable PhysioNet requirements; and StreetAware is
evaluated through its released anonymized recording and documented
governance evidence.

The system-facing evaluation additionally examines authorized research
software and bounded local or prototype validation paths. The CAPTURE
evaluation examined an administrative component that configures sensing
deployments. The BBQS uploader and Shepherd ingest
pipeline are evaluated at documented software interfaces; the PACT
uploader preflight is a prototype extension rather than an archive-side
production control. These evaluations do not modify participant data,
clinical records, production credentials, or study authorization
state. We report software capability and implementation evidence
separately from claims of deployment, authorization, or enforcement.

\paragraph{Privacy and Identity Claims}
Our experiments measure bounded properties such as participant
distinguishability, representation recurrence, candidate association,
and within-recording appearance-similarity association. They do not perform
real-world identity verification or attempt to contact or identify
individuals. In particular, the MIMIC analysis lacks cross-release
patient-identity ground truth and therefore does not establish
same-patient linkage or re-identification. StreetAware uses
within-recording pseudo-tracks rather than persistent or named
identities, and its face-channel experiment measures within-recording
appearance-similarity association rather than facial identity recovery.

\paragraph{Governance Interpretation and Dual Use}
\name\ separates empirical observations from governance claims, so
successful inference alone is not treated as Contradictory. Because some evaluated techniques
could otherwise be interpreted as general-purpose re-identification
methods, we report bounded evaluation protocols rather than operational
identity-recovery pipelines and make identity claims only where
supported by appropriate ground truth.

\paragraph{Data Access and Reproducibility}
The evaluation uses datasets subject to distinct access and
redistribution conditions. MIMIC-III/MIMIC-IV are available through
PhysioNet under applicable credentialing requirements, StreetAware data
are obtained from their original provider, and STM-HAR materials remain
subject to the governing IRB protocol. We do not redistribute restricted
participant-level or clinical data. Reproducibility is therefore
constrained by independent dataset access, applicable permissions, and
the availability of the original evaluation inputs and software
components.

\paragraph{Limitations}
\name\ does not determine legal compliance, infer unstated obligations,
or establish that a Consistent interface is privacy-safe.
Likewise, the absence of a measured finding under the evaluated
adversaries does not establish the absence of other privacy risks.
Additional interpretation or source clarification remains necessary when
governance evidence is ambiguous, incomplete, or context dependent.

\section{Generative AI Usage Considerations}
\label{sec:generative_ai_usage}

Generative AI was used in our research methodology and manuscript
preparation. In this study, large language models were used to support
the interpretation of documented privacy-governance statements and the
evaluation of evidence-based governance relationships: coding agents
assisted the investigators in constructing source-linked governance
records, an automated LLM extraction pipeline produced alternative
representations, and controlled evaluation sessions examined the
consistency and stability of governance interpretations under different
assumptions. The PACT implementation itself does not invoke a language
model or generate governance relations automatically.

Model-generated outputs are not treated as authoritative governance
requirements or independent ground truth. We
assess decision consistency and acknowledge limitations related to model
variability, interpretation uncertainty, and reproducibility. Human
oversight remains necessary when governance evidence is incomplete or
context-dependent.

Generative AI was also used for editorial assistance during manuscript
preparation. All generated content was reviewed by the authors, who
remain fully responsible for the accuracy, originality, citations,
methodology, and conclusions of the paper.

\end{document}